\documentclass[twocolumn,pra,showpacs]{revtex4}
\usepackage{amssymb}
\usepackage{amsmath}
\usepackage[dvips]{graphicx}

\newcommand{\vect}[1]{\mathbf{#1}}
\newcommand{\tens}[1]{\overleftrightarrow{#1}}

\begin{document}

\title{Orientation-dependent Casimir force arising from highly anisotropic crystals: application to Bi$_2$Sr$_2$CaCu$_2$O$_{8+\delta}$}
\author{Mark B. Romanowsky}
\affiliation{Department of Physics, Harvard University, Cambridge, MA 02138, USA}
\author{Federico Capasso}
\email{capasso@seas.harvard.edu}
\affiliation{School of Engineering and Applied Sciences, Harvard University, Cambridge, MA 02138, USA}
\pacs{12.20.Ds, 74.25.Nf, 74.72.Hs}
\date{September 18, 2008}

\begin{abstract}

We calculate the Casimir interaction between parallel planar crystals of Au and the anisotropic cuprate superconductor Bi$_2$Sr$_2$CaCu$_2$O$_{8+\delta}$ (BSCCO), with BSCCO's optical axis either parallel or perpendicular to the crystal surface, using suitable generalizations of the Lifshitz theory. We find that the strong anisotropy of the BSCCO permittivity gives rise to a difference in the Casimir force between the two orientations of the optical axis, which depends on distance and is of order 10-20\% at the experimentally accessible separations 10 to 5000 nm.
\end{abstract}

\maketitle

The Casimir force, an attraction between conductors that arises from the quantum mechanics of electromagnetism in the vacuum, has been of interest to fundamental physics since its first description by Casimir \cite{Casimir}. Recently, however, interest has grown in ``tailoring'' the Casimir force by adjusting material properties like film reflectivity \cite{HydroMirror}, thickness \cite{thinFilm}, and carrier density \cite{SiCasimir}. Some recent proposals suggest that it might even be possible to reverse the sign of the Casimir force between negative-index materials \cite{Leonhardt} or other materials with unusual electromagnetic properties \cite{Kenneth}, which could be realized in properly designed metamaterials \cite{metamaterials}. Since most, if not all, designs for such metamaterials are anisotropic, it is necessary to consider what effect this anisotropy has on the Casimir interaction.

Here we show that strong anisotropy in the dielectric permittivity alone has a significant influence on the Casimir interaction. We calculate the Casimir force between crystals of Au and the cuprate superconductor Bi$_2$Sr$_2$CaCu$_2$O$_{8+\delta}$ (BSCCO); when the optical axis of the BSCCO crystal is switched between being parallel and perpendicular to the crystal surface, the Casimir force changes by a distance-dependent amount, up to 25\%, at the experimentally accessible separations 10 to 5000 nm. BSCCO is well studied as a high-temperature cuprate superconductor, though superconductivity is not at issue here. Rather, BSCCO is an exemplary material for the present study because it shows great anisotropy over a very wide frequency range, yet is a homogeneous single crystal and can be modeled with its dielectric permittivity without having to consider its microscopic structure, as must be done at high frequencies with artificial metamaterials.

Some relevant theoretical work has been done already. In \cite{Parsegian}, the van der Waals force was calculated between anisotropic plates in the non-retarded limit, which is only applicable to separations of tens of \AA{} or less. In \cite{Mostepanenko}, the Casimir-Polder force was calculated between an atom and a uniaxial crystal surface in one orientation (called the perpendicular cleave in the present work; see below), and in \cite{Mostepanenko2006} the Casimir force between an isotropic surface and a uniaxial surface in the same orientation, both papers dealing primarily with graphite and graphene. Recently, \cite{Rosa} studied specifically the Casimir force between a metal and an anisotropic metamaterial, though again only in one orientation (the perpendicular cleave), and at only one separation. In \cite{BarashRussian,BarashCapasso}, the Casimir-Lifshitz interaction between two uniaxial surfaces was analyzed in a different orientation (called the parallel cleave in the present work) with the goal of calculating a quantum electrodynamic torque, but the relation of anisotropy to the distance dependence of the force was largely unexplored. In the following, we show that in fact this distance dependence can be made to vary significantly by changing the orientation of the crystals. 

A uniaxial crystal is characterized by a dielectric permittivity that has the value $\epsilon_\parallel (\omega)$ for electric fields polarized along the so-called optical axis, and a different value $\epsilon_\perp(\omega)$ for electric fields polarized perpendicular to the optical axis. The permittivity depends on the angular frequency $\omega$ of the field, but the direction of the optical axis is the same for all frequencies in a uniaxial crystal. Such crystals generally are birefringent, and waves with the same frequency in the same direction will have a wave number (or wavelength) that depends on their polarization. Therefore if a cavity is composed of uniaxial crystal surfaces, the electromagnetic normal modes will have different dispersion relations for the two principal polarizations, which affects the zero-point energy of the modes, and in turn the Casimir interaction.

\begin{figure}
	\centering
		\includegraphics[width=0.48\textwidth]{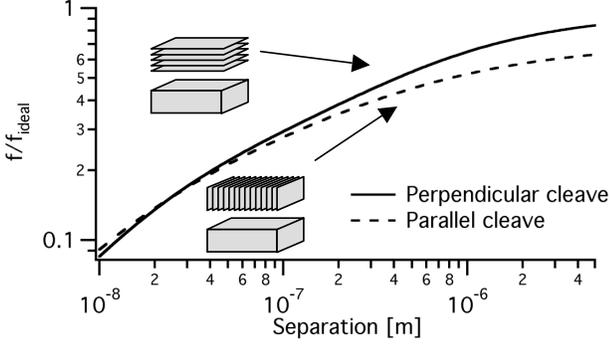}
		\caption{The Casimir force between Au and BSCCO depends on the orientation of the BSCCO optical axis. Solid curve is $f_{perp}$, the force per unit area for the perpendicular cleave, dashed curve is $f_{par}$ for the parallel cleave, both normalized to the result for ideal metal plates $f_{ideal}=-\pi^2\hbar c/240d^4$. The insets show the orientation of the conducting copper oxide planes in BSCCO. The optical axis is perpendicular to these planes.}
	\label{fig:normForce}
\end{figure}

In the present paper we calculate the Casimir force between uniaxial crystal surfaces in two orientations (shown schematically in Fig.~\ref{fig:normForce}). The first orientation, which we call the \emph{perpendicular cleave}, has the optical axis perpendicular to the surfaces. The second, which we call the \emph{parallel cleave}, has the optical axis in the plane of the surfaces. (We note that the parallel cleave can give rise to a QED torque as shown in \cite{BarashRussian,BarashCapasso}.)

We consider first the perpendicular cleave. (A similar derivation is used in \cite{Rosa}.) Using Eqs.~86 and 90 from \cite{Lambrecht}, we calculate the Casimir force per unit area $f(d)$ for parallel plates, made of media 1 and 2, separated by a distance $d$ of vacuum:
\begin{equation}
f(d)=2\hbar \sum_{p} \int \frac{d^2\vect{k}}{4\pi^2} \int_0^\infty \frac{d\xi}{2\pi} \kappa_0 \frac{r_{1,\vect{k}}^{p} r_{2,\vect{k}}^{p} e^{-2\kappa_0 d}}{ 1- r_{1,\vect{k}}^{p} r_{2,\vect{k}}^{p} e^{-2\kappa_0 d}}
\end{equation}
where $r_{j,\vect{k}}^p = r_{j,\vect{k}}^p (i\xi)$ is the reflection amplitude at an interface between vacuum and material $j$ for a wave whose transverse component of the wave vector is $\vect{k}$, imaginary frequency is $i\xi$, and polarization $p\in\{\textrm{TE, TM}\}$, and $\kappa_0 = \sqrt{k^2 + \xi^2/c^2}$. The reflection amplitudes are well defined for the perpendicular cleave because TE and TM are the principal polarizations for waves traveling in the uniaxial crystals (but not for other orientations of the optical axes). If the crystal-vacuum interfaces lie in the $xy$ plane, the optical axis  points in the $z$ direction and the permittivity in material $j$ is the tensor
\begin{equation}
\tens{\epsilon_j} = 
\left(
\begin{array}{ccc}
\epsilon_{j\perp} & 0 & 0\\
0 & \epsilon_{j\perp} & 0\\
0 & 0 & \epsilon_{j\parallel}
 \end{array} \right).
\end{equation}
(We suppress the $\omega$ dependence of all permittivities for readability.) The TE and TM reflection amplitudes are \cite{TETM}
\begin{equation}
r_{j,\vect{k}}^{TE} (\omega)=\frac{\kappa_0 - \kappa_j}{\kappa_0 + \kappa_j},\ r_{j,\vect{k}}^{TM}(\omega)=\frac{\epsilon_{j\perp}\kappa_0 - \lambda_j}{\epsilon_{j\perp}\kappa_0 + \lambda_j}
\end{equation}
where $\kappa_0 = \sqrt{k^2-\omega^2/c^2},$ $\kappa_j = \sqrt{k^2 -\epsilon_{j\perp}\omega^2/c^2}$, and $\lambda_j = \sqrt{k^2 \epsilon_{j\perp} / \epsilon_{j\parallel} - \epsilon_{j\perp} \omega^2/c^2}$.
Thus for this cleave
\begin{widetext}
\begin{equation}
f_{perp}(d) = 2\hbar \int \frac{d^2 \vect{k}}{4\pi^2} \int_0^\infty \frac{d\xi}{2\pi}
\kappa_0 \left\{ 
\left[ \frac{(\kappa_0 + \kappa_1)(\kappa_0 + \kappa_2)}{(\kappa_0 - \kappa_1)(\kappa_0 - \kappa_2)} e^{2\kappa_0 d} -1 \right]^{-1} + 
\left[ \frac{(\epsilon_{1\perp} \kappa_0 + \lambda_1)(\epsilon_{2\perp} \kappa_0 + \lambda_2)}{(\epsilon_{1\perp}\kappa_0 - \lambda_1)(\epsilon_{2\perp} \kappa_0 - \lambda_2)} e^{2\kappa_0 d} -1 \right]^{-1} \right\}
\end{equation}
\end{widetext}
where we have made the substitution $\omega = i\xi$ \cite{KKnote}, so $\kappa_0 = \sqrt{k^2 + \xi^2/c^2}$, 
$\kappa_j = \sqrt{k^2+\epsilon_{j\perp} \xi^2/c^2}$, 
$\lambda_j = \sqrt{k^2 \epsilon_{j\perp} / \epsilon_{j\parallel} + \epsilon_{j\perp}\xi^2/c^2}$, and the permittivities $\epsilon_j=\epsilon_j(i\xi)$ are evaluated at imaginary frequency $i\xi$, for materials $j=1,2$. For isotropic materials, $\epsilon_{j\perp} = \epsilon_{j\parallel}$ and $\lambda_j = \kappa_j$, and we recover the standard Lifshitz formula \cite{Lifshitz}.

In the parallel cleave orientation, we cannot use the same method because the principal polarizations in the uniaxial crystals do not coincide with the TE and TM polarizations. Therefore, the normal mode frequencies must be calculated for the entire cavity at once, keeping track of all three components of both electric and magnetic fields at both interfaces. This straightforward but quite cumbersome calculation is performed in \cite{BarashRussian} (and the result is reported also in \cite{BarashCapasso}), yielding the free energy as a function of temperature and relative angle between the optical axes of the two materials. In the case at hand, we choose BSCCO as material 1, Au as material 2, and vacuum as the intervening medium 3, and also take the zero-temperature limit of the expression. (For the more general and more unwieldy form with anisotropic materials 1 and 2, arbitrary intervening medium, and nonzero temperature, see Eqs.~2.1-2.11 of \cite{BarashCapasso}.) In this case the permittivity tensors are
\begin{equation}
\tens{\epsilon_1} = 
\left(
\begin{array}{ccc}
\epsilon_{1\parallel} & 0 & 0\\
0 & \epsilon_{1\perp} & 0\\
0 & 0 & \epsilon_{1\perp}
 \end{array} \right), \quad
\tens{\epsilon_2} = 
\epsilon_2  \left(
\begin{array}{ccc}
1 & 0 & 0\\
0 & 1 & 0\\
0 & 0 & 1
\end{array} \right) .
\end{equation}
With medium 2 and 3 isotropic, we obtain a greatly simplified expression for the Casimir potential energy per unit area $u_{par}(d)$:
\begin{equation}
u_{par}(d) = \hbar \int_0^\infty \frac{d\xi}{2\pi}  \int_0^\infty \frac {k dk}{4\pi^2} \int_0^{2\pi}d\phi \ln \left[\frac{D(d)}{D_\infty} \right]
\end{equation}
where
\begin{equation}
\begin{split}
D(d) &=  \left[ (\kappa_1 + \kappa_0)(\kappa_2 + \kappa_0) -(\kappa_1 - \kappa_0)(\kappa_2 - \kappa_0)e^{-2\kappa_0 d} \right]
\\
 & \times\left[ (\kappa_1 + \epsilon_{1\perp}\kappa_0) (\kappa_2 + \epsilon_2\kappa_0) - (\kappa_1 - \epsilon_{1\perp}\kappa_0)\right.
\\
 & \times \left. (\kappa_2 - \epsilon_2\kappa_0) e^{-2\kappa_0 d} \right] 
 -\frac{(\tilde\kappa_1 - \kappa_1)\epsilon_{1\perp}}{\kappa_1^2 - k^2 \sin^2 \phi} 
\biggl\{ (k^2\sin^2\phi 
\\
& - \kappa_1\kappa_0)(\kappa_2+\epsilon_2\kappa_0)(\kappa_2+\kappa_0)(\kappa_1+\kappa_0) 
+2(\epsilon_2 - 1)
\\
& \times \left[ k^2\sin^2\phi (k^2\kappa_1 - \kappa_2\kappa_0^2) + \kappa_1\kappa_0^2(k^2 - 2k^2\sin^2\phi \right.
\\
& \left. + \kappa_1\kappa_2)\right]e^{-2\kappa_0 d} 
+(k^2 \sin^2\phi+\kappa_1\kappa_0)(\kappa_2-\epsilon_2\kappa_0)
\\  
& \times (\kappa_1-\kappa_0)(\kappa_2-\kappa_0)e^{-4\kappa_0 d}\biggr\}
\end{split}\end{equation}
\begin{equation}\begin{split}
D_\infty = &\quad (\kappa_1+\kappa_0)(\kappa_2+\kappa_0)(\kappa_1+\epsilon_{1\perp}\kappa_0)(\kappa_2+\epsilon_2\kappa_0)\\
&-\frac{(\tilde\kappa_1 - \kappa_1)\epsilon_{1\perp}}{\kappa_1^2 - k^2 \sin^2\phi} 
(k^2\sin^2\phi - \kappa_1\kappa_0)(\kappa_2 +\epsilon_2\kappa_0)
\\
& \times (\kappa_2+\kappa_0)(\kappa_1+\kappa_0)
\end{split}\end{equation}
\begin{equation}\begin{split}
\kappa_1 &= \sqrt{k^2+\epsilon_{1\perp}\xi^2/c^2},\ \kappa_2 = \sqrt{k^2+\epsilon_2\xi^2/c^2}, \\
 \tilde\kappa_1 &= \sqrt{k^2 - (1-\epsilon_{1\parallel}/\epsilon_{1\perp}) k^2\cos^2\phi + \epsilon_{1\parallel}\xi^2/c^2}, \\
 \kappa_0 &= \sqrt{k^2+\xi^2/c^2}.
\end{split}\end{equation}
As before, $\vect{k}=(k\cos\phi, k\sin\phi)$ is the transverse component of the wave vector with $k,\phi$ its polar coordinates, and the dielectric functions $\epsilon_j = \epsilon_j(i\xi)$ are evaluated at imaginary frequency $i\xi$. The Casimir force per unit area is $f_{par}(d) = -\partial u_{par}/ \partial d$. If material 1 is isotropic, as well as material 2, $\epsilon_{1\perp}=\epsilon_{1\parallel}$ and we recover the standard Lifshitz formula. 

\begin{figure}
  \centering
    \includegraphics[width=0.46\textwidth]{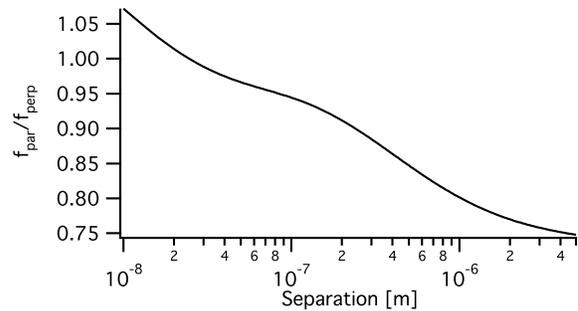}
    \caption{The ratio of Casimir forces $f_{par}/f_{perp}$ between BSCCO and Au, for the two orientations of the optical axis. The two forces are most different at large separations, where long wavelength permittivity dominates and the anisotropy of BSCCO is most pronounced.}
  \label{fig:ratio}
\end{figure}

We now evaluate the Casimir force for our example case, taking the cuprate superconductor Bi$_2$Sr$_2$CaCu$_2$O$_{8+\delta}$ (BSCCO) for material 1 and Au for material 2. We assume the BSCCO to be optimally doped, and consider its normal-state permittivity. Like other cuprate superconductors, BSCCO is composed of layers of conductive copper oxide planes separated by insulating oxides; the conductivity is much higher in the copper oxide plane than perpendicular to it, and the permittivity is likewise anisotropic. (There is also a much smaller anisotropy between the two principal axes within the copper oxide planes \cite{Quijada}, which we neglect.) Due to the extreme anisotropy of BSCCO, the Casimir force differs in the two orientations by up to 10\% for $10\textrm{ nm}<d<250\textrm{ nm}$, and by up to 25\% at larger separations, as shown in Fig.~\ref{fig:normForce}; forces are normalized to the ideal metal result \cite{Casimir} $f_{ideal}=-\pi^2\hbar c/240d^4$. The ratio $f_{par}/f_{perp}$ is shown in Fig.~\ref{fig:ratio}. For $d>25\textrm{ nm}$, the force is stronger for the perpendicular cleave than the parallel cleave. For $d<25\textrm{ nm}$, the reverse is true.

\begin{figure}
  \centering
    \includegraphics[width=0.46\textwidth]{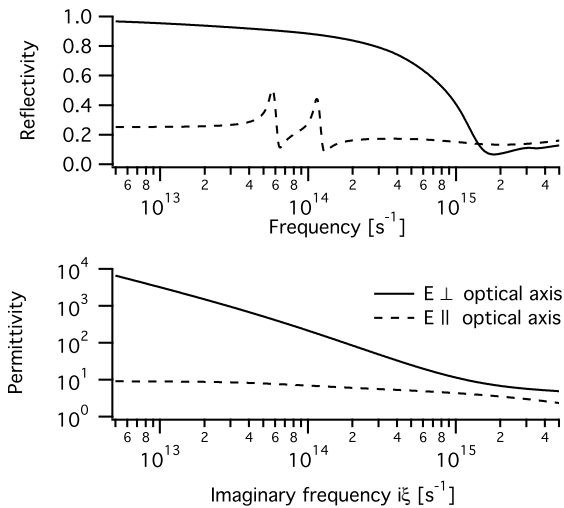}
    \caption{Optical properties of BSCCO. Waves with electric field perpendicular to the optical axis meet with a very conductive and reflective material, compared with waves polarized parallel to the optical axis. Top panel: reflectivity at normal incidence vs. real frequency for the two polarizations. Bottom panel: permittivity vs. imaginary frequency for the two polarizations.}
  \label{fig:ReflEpsLayout}
\end{figure}

To model the permittivity function of BSCCO, we use multi-oscillator models where
\begin{equation}
\label{model1}
\epsilon_{j}(i\xi) = 1+ \sum_{m=1}^{N_j} \frac{A_{m,j}^2}{\xi^2+\xi\gamma_{m,j} + \omega_{m,j}^2}.
\end{equation}
We model the permittivity of Au using a Drude component plus two interband transitions in the critical point transition model \cite{AuAnalytic} (since Au is not well described by a small number of Lorentz oscillators):
\begin{equation}
\label{model2}
\epsilon_{Au}(i\xi) = \epsilon_\infty +\frac{\omega_p^2}{\xi^2+\gamma_0\xi} +\sum_{m=1}^{2} \frac{C_m (\omega_m + \gamma_m +\xi)}{\omega_m^2 + (\xi + \gamma_m)^2}.
\end{equation}
The parameters are shown in Table \ref{paramTable} (based on \cite{AuAnalytic} for Au, on \cite{Quijada} for $\epsilon_\perp$ of BSCCO, and on our fits to reflectivity \cite{Tajima} and permittivity data \cite{TajimaUnpublished} for $\epsilon_\parallel$ of BSCCO). The anisotropy of BSCCO in our model is shown in Fig.~\ref{fig:ReflEpsLayout}: the top panel shows a much higher reflectivity (at normal incidence) for waves polarized perpendicular to the optical axis, compared with waves polarized parallel to the optical axis, over a wide range of frequencies; the bottom panel shows the same information presented as permittivity versus imaginary frequency.

To perform the computations, we use \textsc{Mathematica} 6 to numerically integrate the expressions for $f_{perp}$ and $f_{par}$. To control a spurious contribution from roundoff error, we include an upper cutoff in the integration for the transverse wave vector and frequency of $10^9\textrm{ m}^{-1}$ and $c\times 10^9\textrm{ m}^{-1}$, respectively. Increasing the cutoffs by a factor of 10 changes our results insignificantly, by less than 1 part in $10^6$, at all separations.

\begin{table}
\begin{ruledtabular}
\begin{tabular}{llll}
BSCCO & $\omega_1 = 0$ &  $A_1=1.66[15]$ & $\gamma_1=8.10[13]$ \\
$\epsilon_\perp$ & $\omega_2 = 1.48[14]$ & $A_2=1.95[15]$ & $\gamma_2=4.24[14]  $\\
& $\omega_3 = 8.06[14]$& $A_3=2.10[15]$& $\gamma_3=1.47[15] $\\
& $\omega_4 = 3.26[15]$& $A_4=1.32[15]$& $\gamma_4=1.18[15] $\\
& $\omega_5 = 1.88[16]$& $A_5=3.62[16]$& $\gamma_5=1.88[14] $\\
\hline
BSCCO & $\omega_1 = 5.65[13]$& $A_1=7.35[13]$& $\gamma_1=4.71[12] $\\
$\epsilon_\parallel$ & $\omega_2 =1.13[14]$ & $A_2=1.22[14]$& $\gamma_2=9.42[12] $\\
& $\omega_3 = 1.22[15]$& $A_3=1.69[15]$& $\gamma_3=2.26[15] $\\
& $\omega_4 = 6.22[15]$ & $A_4= 1.13[16]$ & $\gamma_4=7.53[15] $\\
\hline
Au & $\epsilon_\infty=1.54$& $\omega_p=1.32[16]$& $\gamma_0=1.30[14]$\\
& $\omega_1=4.01[15]$& $C_1=7.20[15]$& $\gamma_1=9.92[14] $\\
& $\omega_2=5.80[15]$& $C_2=9.02[15]$& $\gamma_2=1.78[15] $\\
\end{tabular}
\end{ruledtabular}
\caption{Parameters used in models of the permittivities of  BSCCO and Au. Resonance frequencies $\omega_m$, widths $\gamma_m$, and strengths $A_m$ and $C_m$ are in rad/s, and $\epsilon_\infty$ is unitless. Powers of 10 are bracketed, e.g.~$2[4]=2\times 10^4$. See Eqs.~\ref{model1} and \ref{model2}.\label{paramTable}}
\end{table}

The difference in force between the two orientations follows our expectations, at least for $d>25\textrm{ nm}$, if we imagine BSCCO's copper oxide planes as a stack of partially reflective mirrors. In the perpendicular cleave, these ``mirrors'' are parallel to the crystal surface and, with the Au surface, constitute a confining Fabry-Perot-like cavity for all modes. In the parallel cleave, the ``mirrors'' are perpendicular to the crystal surface and resemble a linear polarizer, and effectively confine modes of only one polarization. We should expect the first case to have a stronger Casimir force than the second, and this indeed happens. At very small separations, the high frequency (visible and UV) optical properties become increasingly relevant, and at these frequencies the reflectivities for the different polarizations are similar. The analogy to mirrors therefore breaks down at very small separations.

The calculation of the Casimir force depends on the permittivity over all frequencies, and the incomplete knowledge of the permittivity function is a major source of uncertainty. Even sample-to-sample variations in the optical properties of nominally the same material can change the force by 5\% or more \cite{LambrechtVariation}. This is the most significant uncertainty in our calculation for Au and BSCCO. Since the reflectivity of BSCCO over the reported frequency range, 30-30000 cm$^{-1}$, is quite different from zero for both polarizations \cite{Tajima}, we can infer that there are resonances at higher frequency, but cannot know for certain their frequency or number based on the data published so far.

For this reason, we add to $\epsilon_\perp(\omega)$ a single oscillator term $A^2_{HF}/(\omega_{HF}^2 -\omega^2 -i\omega\gamma_{HF})$ with center frequency $\omega_{HF}$ much higher than the highest frequency for which we have data. This contributes a constant term $A^2_{HF}/\omega_{HF}^2$ for all $\omega \ll \omega_{HF}$. We chose $\omega_{HF} = 1.88\times 10^{16}\textrm{ rad/s}$, corresponding to a wavelength of 100 nm. A lower $\omega_{HF}$ would conflict with existing optical data. Increasing $\omega_{HF}$ (while holding $A^2_{HF}/\omega_{HF}^2$ constant) does not have a significant effect on our conclusions: raising it by a factor of 10 changes the the ratio $f_{perp}/f_{par}$ by less than 1\% at separations $d<100\textrm{ nm}$ (though the overall magntitude of $f_{perp}$ and $f_{par}$ changes by a few percent), and has even less effect at larger separations. Further increasing $\omega_{HF}$ does not further change the Casimir interaction.

To choose our model for BSCCO's $\epsilon_\parallel (\omega)$, we used the published reflectivity data \cite{Tajima} as well as unpublished permittivity data covering 0.1 to 4000 cm$^{-1}$ \cite{TajimaUnpublished}. We obtained a satisfactory fit to the optical data using a highest resonance at 33,000 cm$^{-1}$.

Regardless of these theoretical uncertainties, this calculation serves as a proof of principle that realistic material systems can show a significant change in Casimir force when their optical axis orientation is changed, even if the calculation is not quantitatively exact for the Au/BSCCO system. In any case, we expect our conclusions to hold qualitatively for this system, as they stem from the established anisotropic permittivity of BSCCO in the dc to infrared frequency range.

In conclusion, we have shown that strong anisotropy in the permittivity of a uniaxial crystal can give rise to a Casimir force that depends significantly on the orientation of the optical axis. We find that the Casimir force between Au and BSCCO plates varies by an amount of order 10-20 \% when the optical axis of BSCCO lies perpendicular versus parallel to the crystal surface. We expect a similar effect for other materials with similar anisotropy. Such an effect represents a way to vary the strength of the Casimir force between two given materials merely by changing their relative orientation. This adds another to the small number of ``handles'' on the Casimir force, which is usually determined only by material properties and the distance between the plates.

\begin{acknowledgments}

We are grateful to S. Tajima for the use of unpublished data on the optical permittivity of BSCCO, and to J.N. Munday for discussions. M.B.R. acknowledges support from the NDSEG Fellowship.

\end{acknowledgments}

\end{document}